\documentclass[aps,prx,superscriptaddress,longbibliography,twocolumn,10pt]{revtex4-2}
\usepackage{graphicx}
\usepackage{amsmath,amssymb,mathtools,bbm,bbold,braket,hyperref,xcolor,multirow,setspace}
\usepackage[utf8]{inputenc}
\usepackage[english]{babel}
\usepackage[margin=0.63in]{geometry}
\usepackage{pgf}
\hypersetup{colorlinks=true, linkcolor=blue, citecolor=blue, urlcolor=blue}

\DeclareMathOperator{\Res}{Res}

\newcommand{\bea}{\begin{eqnarray}}
\newcommand{\eea}{\end{eqnarray}}

\begin{document}

\title{A Compact Analytical Solution of the Dicke Superradiance Master Equation via Residue Calculus}

\author{Raphael Holzinger}
\email{raphael\_holzinger@fas.harvard.edu}
\affiliation{Department of Physics, Harvard University, Cambridge, Massachusetts 02138, USA}
\affiliation{Institute for Theoretical Physics, University of Innsbruck, Technikerstra\ss e 21a, 6020 Innsbruck, Austria}

\author{Claudiu Genes}
\email{claudiu.genes@physik.tu-darmstadt.de}
\affiliation{TU Darmstadt, Institute for Applied Physics, Hochschulstra\ss e 4A, D-64289 Darmstadt, Germany}
\affiliation{Max Planck Institute for the Science of Light, Staudtstra\ss e 2, D-91058 Erlangen, Germany}

\begin{abstract}
We revisit the Dicke superradiance problem, where an ensemble of $N$ identical two-level systems undergoes collective spontaneous decay. While an exact analytical solution has been known since 1977, its algebraic complexity has hindered practical use. Here we present a compact, closed-form solution that expresses the dynamics as a finite sum over residues or, equivalently, a complex contour integral. The method yields explicit populations of all Dicke states at arbitrary times and system sizes, and generalizes naturally to arbitrary initial conditions. Our formulation is computationally efficient and offers structural insights into the role of spectral degeneracies and Lindbladian eigenmodes in collective decay.
\end{abstract}

\maketitle


Dicke superradiance is a collective quantum phenomenon in which an ensemble of \( N \) identical two-level systems (often modeled as spin-1/2 particles or two-level atoms), all initially prepared in their excited states, spontaneously emits radiation in a short, intense burst. This effect is fundamentally different from the independent emission of uncorrelated atoms: the radiated intensity scales quadratically with the number of emitters (\( \propto N^2 \)) rather than linearly, reflecting the highly correlated nature of the emission process. The key physical mechanism behind this enhancement is the constructive interference of the quantum dipole amplitudes of the emitters, which leads to a build-up of a macroscopic polarization and an accelerated emission of photons. The phenomenon was first described by R.H. Dicke in 1954~\cite{Dicke_originalpaper}, who introduced a symmetric basis of collective atomic states—now known as Dicke states—to account for these correlations in a fully quantum description.\\
\indent An exact analytical solution to the time evolution of this process, starting from the fully inverted (i.e., all-excited) initial state, was later provided by C.T. Lee in 1977~\cite{Tsung1977Part1}. This work established the formal solvability of the problem using a recursive method within the symmetric subspace of the full Hilbert space. Shortly thereafter, the same author extended the analysis to more general initial conditions, including partial excitation~\cite{Tsung1977Part2}, thereby significantly broadening the scope of analytical accessibility. These results, though rigorous, involve expressions that grow rapidly in complexity and are not easily handled for large system sizes.\\
\indent Additional analytical insights were obtained through alternative approaches developed in the 1980s. In particular, Bethe ansatz techniques and field-theoretic formalisms were employed to provide a different perspective on the dynamics of Dicke superradiance. These methods, notably those introduced in Refs.~\cite{Rupasov1984,Yudson1985}, treat the ensemble of emitters as an integrable quantum system coupled to a common radiation field. The use of Bethe ansatz allows one to construct exact eigenstates of the many-body Hamiltonian and to study the emission process in terms of scattering amplitudes of virtual photons. In this formulation, the time evolution of the system is encoded in correlation functions that can, in principle, be computed exactly, providing access to the full non-equilibrium dynamics.\\
\indent Such techniques extend the reach of analytical methods beyond the symmetric Dicke manifold, offering insights into systems with spatial or spectral inhomogeneities, or into regimes where the Markovian approximation breaks down. Although mathematically demanding, these approaches have deepened our understanding of superradiance as a manifestation of collective behavior in open quantum systems, and they establish connections with broader topics in integrable systems, non-equilibrium quantum field theory, and many-body quantum optics. Together, these contributions form a rich theoretical framework in which superradiance appears not only as a textbook example of cooperative emission, but also as a touchstone for exploring the interplay of coherence, entanglement, and dissipation in quantum many-body physics.\\
\indent The central theoretical task for Dicke superradiance is to compute the time evolution of the system's density operator \( \rho(t) \) under collective decay. In the interaction picture, the dynamics are governed by a Lindblad master equation of the form
\begin{equation}
\dot{\rho}(t) = \mathcal{L}[\rho] = \Gamma \left[ S \rho S^\dagger - \frac{1}{2} \left( S^\dagger S \rho + \rho S^\dagger S \right) \right],
\label{master}
\end{equation}
where \( \Gamma \) denotes the collective decay rate and \( S \) is the collective lowering operator.\\
\indent This operator acts within the fully symmetric subspace of the total Hilbert space, spanned by Dicke states \( \ket{m} \) with excitation numbers \( m = 0, 1, \dots, N \). The ladder structure is defined by
\[
S \ket{m} = \sqrt{h_m} \ket{m-1}, \quad S^\dagger \ket{m} = \sqrt{h_{m+1}} \ket{m+1},
\]
with \( h_m = m(N+1 - m) \). For even \( N \), the spectrum is symmetric about \( m = N/2 \), and \( h_m = h_{\bar{m}} \) with \( \bar{m} = N+1 - m \). For odd \( N \), this degeneracy is broken at the central state.\\
\indent Despite the existence of formal analytical solutions~\cite{Tsung1977Part1,Tsung1977Part2}, their algebraic complexity renders them impractical for computation, especially when compared to direct numerical integration of the master equation. As a result, recent attention has shifted toward approximate treatments valid in the thermodynamic limit \( N \to \infty \); see, for example, Refs.~\cite{approx1,approx2,approx3,approx4,molmer2021,Cirac2022LargeNLimit}. A detailed early review is found in Ref.~\cite{gross_haroche}.

While Dicke's original formulation considered emitters at zero separation—an unphysical limit—modern realizations are achieved in cavity and circuit quantum electrodynamics. In particular, atoms in a lossy optical cavity (with coherent coupling \( g \) and cavity loss \( \kappa \)) undergo effective collective decay with rate \( \Gamma = g^2/\kappa \), following adiabatic elimination of the cavity field~\cite{bohnet2012asteadystate}. Provided spontaneous emission is negligible, such systems realize the model of Eq.~\eqref{master}. Similar implementations exist in waveguide QED and superconducting circuit platforms~\cite{Superradiance2016Lambert,zanner2022coherent}, where collectively enhanced emission has been experimentally demonstrated.
\\

\noindent \textbf{Solution} -- We now present an exact analytical solution to the Dicke superradiance problem, written in a compact form that is both transparent and numerically tractable. We consider the density matrix in the symmetric Dicke basis, where the state of the system at time \( t \) is fully described by the diagonal elements
\[
\rho(t) = \sum_{m=0}^N \rho_m(t) \ket{m}\bra{m},
\]
as all off-diagonal elements, or coherence, are always zero. Each population \( \rho_m(t) \) evolves according to
\begin{equation}
\rho_m(t) = \sum_{j = j_m}^N \Res\left[f_m(z,t)\right] \Big|_{z = h_j},
\end{equation}
where the lower summation limit \( j_m \) depends on the parity of \( N \): for even \( N \), we define \( j_m = \max\{m, N/2 + 1\} \), and for odd \( N \), \( j_m = \max\{m, (N+1)/2\} \). The function \( f_m(z,t) \), defined below, has simple or double poles located at the values \( z = h_j \), which reflect the decay channels of the system.

For even \( N \), the explicit expression for \( \rho_m(t) \) reads:
\begin{widetext}
\begin{equation} \label{solution}
\rho_m(t) =
\sum_{j = m}^N \left[ f_m(z,t)(z - h_j) \right]_{z = h_j} \, \theta_{m > \frac{N}{2}} 
+
\left\{ 
\sum_{j = \bar{m} + 1}^N \left[ f_m(z,t)(z - h_j) \right]_{z = h_j}
+ 
\sum_{j = N/2 + 1}^{\bar{m}} \frac{d}{dz} \left[ f_m(z,t)(z - h_j)^2 \right]_{z = h_j}
\right\} \theta_{m \leq \frac{N}{2}},
\end{equation}
\end{widetext}
where \( \bar{m} = N + 1 - m \) and the notation \( [\cdot]_{z = h_j} \) indicates that the expression is evaluated at \( z = h_j \). The Heaviside functions \( \theta_{m > N/2} \) and \( \theta_{m \leq N/2} \) are defined to be 1 if the condition is satisfied and 0 otherwise. These functions distinguish between the non-degenerate region above the equator (where \( h_m \ne h_{\bar{m}} \)) and the degenerate region below it (where \( h_m = h_{\bar{m}} \)).

The generating function whose residues appear in the solution is given by
\begin{equation} \label{eq:function}
f_m(z,t) = (-1)^m \frac{h_N \cdots h_{m+1}}{(z - h_N) \cdots (z - h_m)} e^{-z \Gamma t}.
\end{equation}
This function captures the structure of the cascade through Dicke states during the collective decay. Each pole \( z = h_j \) corresponds to an eigenvalue of the Lindblad superoperator \( \mathcal{L} \), and the full solution can be understood as a linear combination of the associated exponential decay terms \( e^{-h_j \Gamma t} \), for \( j = 1, \ldots, N \).

\begin{figure}[b]
    \centering   
\includegraphics[width=0.95\columnwidth]{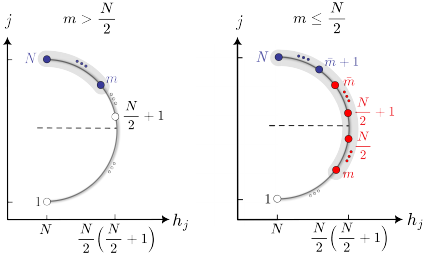}
    \caption{Left: Contributions to the solution in Eq.~\eqref{solution} above the equator adding only simple poles marked in blue. Right: Separation of simple pole contributions in blue and double poles contributions ($h_j=h_{\bar{j}}$) marked in red, for even $N$.}
    \label{fig:1}
\end{figure}

As illustrated in Fig.~\ref{fig:1}, for \( m > N/2 \), all contributing poles are simple and non-degenerate. For \( m \leq N/2 \), the presence of degeneracies requires evaluating second-order residues, which are handled via the derivative terms in Eq.~\eqref{solution}. Despite the complexity of the expression, the final result is readily computable for large \( N \) and provides exact time-dependent populations without requiring direct numerical integration of the master equation.\\

\noindent \textbf{Derivation} --  
We now describe the analytical derivation leading to the exact solution introduced above. Our aim is to compute the time-dependent populations \( \rho_m(t) = \langle m | \rho(t) | m \rangle \), that is, the diagonal elements of the density matrix in the Dicke basis.

To begin, we express the time-evolved density operator using a power series expansion in the Lindbladian superoperator \( \mathcal{L} \). Applied to the initial condition \( \rho(0) = \ket{N} \bra{N} \), the evolution reads:
\begin{equation} \label{power-series}
\rho_m(t) = \langle m | e^{\mathcal{L} t} \rho(0) | m \rangle = \sum_{j=0}^\infty \frac{(\Gamma t)^j}{j!} \rho_m^{(j)},
\end{equation}
where each coefficient \( \rho_m^{(j)} \) corresponds to the action of the \( j \)-th power of the Lindbladian on the initial state:
\[
\rho_m^{(j)} = \langle m | \mathcal{L}^j[\rho(0)] | m \rangle.
\]

To compute these terms, we examine the recursive structure generated by applying \( \mathcal{L} \) to projectors onto Dicke states:
\begin{equation}
\mathcal{L}[ \ket{m} \bra{m} ] = -h_m \ket{m}\bra{m} + h_m \ket{m-1} \bra{m-1},
\end{equation}
where \( h_m = m(N + 1 - m) \) are the familiar Dicke ladder factors. This identity reflects the fact that the system may either remain in the current Dicke state (with weight \( -h_m \)) or decay to the next lower excitation level (with weight \( h_m \)).

Since the bottom state \( \ket{0} \) has no further decay, we find \( \mathcal{L}[ \ket{0} \bra{0} ] = 0 \). Using this recursive structure, we can compute higher-order corrections \( \rho_m^{(j)} \) by applying the following recurrence relation:
\begin{equation}
\rho_{m-1}^{(j+1)} = -h_{m-1} \rho_{m-1}^{(j)} + h_m \rho_m^{(j)}.
\end{equation}
This recurrence allows us to construct the full set of coefficients starting from the initial condition \( \rho_N^{(0)} = 1 \) (i.e., all population initially in the fully excited state). From this, we find:
\[
\rho_N^{(j)} = (-h_N)^j \quad \Rightarrow \quad \rho_N(t) = e^{-h_N \Gamma t} = e^{-N \Gamma t}.
\]

This recurrence can be carried out systematically to obtain closed-form expressions for each \( \rho_m(t) \). For levels above the equator (\( m > N/2 \)), the recursive solution results in a compact expression:
\begin{equation} \label{SolPlus}
\rho_m(t) = \sum_{j=m}^N (-1)^m \frac{ h_N \cdots h_{m+1} }{ (h_j - h_N) \cdots 1 \cdots (h_j - h_m) } e^{-h_j \Gamma t},
\end{equation}
where the denominator omits the term \( h_j - h_j \), which is formally replaced by 1. This reflects that each term in the sum arises from a first-order (simple) pole at \( z = h_j \), as no degeneracy occurs in this region.

To clarify this structure, we define the function:
\begin{equation}
f_m(z,t) = \mathcal{P}_m(z) e^{-z \Gamma t},
\end{equation}
with the rational part
\begin{equation}
\mathcal{P}_m(z) = (-1)^m \frac{ h_N \cdots h_{m+1} }{ (z - h_N) \cdots (z - h_m) }.
\end{equation}
The contributions to \( \rho_m(t) \) are then seen as residues of \( f_m(z,t) \) evaluated at the poles \( z = h_j \):
\begin{equation}
\left[ f_m(z,t)(z - h_j) \right]_{z = h_j} = \mathcal{P}_{mj} e^{-h_j \Gamma t},
\end{equation}
with
\begin{equation}
\mathcal{P}_{mj} = (-1)^m \frac{ h_N \cdots h_{m+1} }{ (h_j - h_N) \cdots 1 \cdots (h_j - h_m) }.
\end{equation}

For Dicke states at or below the equator (\( m \leq N/2 \)), a complication arises: the spectrum of the Dicke ladder becomes degenerate due to the symmetry \( h_m = h_{\bar{m}} \), where \( \bar{m} = N + 1 - m \). This degeneracy leads to the appearance of second-order (double) poles in the generating function \( f_m(z,t) \), which must be handled via residue calculus involving derivatives.

These degenerate terms arise in the interval \( j \in [m, \bar{m}] \), and their contributions are computed from second-order residue formulas:
\begin{equation}
\left. \frac{d}{dz} \left[ f_m(z,t)(z - h_j)^2 \right] \right|_{z = h_j} = \mathcal{R}_{mj}(t) e^{-h_j \Gamma t},
\end{equation}
with the prefactor given by
\begin{equation}
\mathcal{R}_{mj}(t) = -\Gamma t \, \tilde{\mathcal{P}}_{mj} - \left. \frac{d}{dz} \tilde{\mathcal{P}}_m(z) \right|_{z = h_j}.
\end{equation}
Here, \( \tilde{\mathcal{P}}_m(z) \) is a modified version of \( \mathcal{P}_m(z) \) in which two terms in the denominator, corresponding to \( j \) and \( \bar{j} \), are removed. The term \( \tilde{\mathcal{P}}_{mj} \) is the result of evaluating this expression at the degenerate value \( z = h_j \):
\begin{equation}
\tilde{\mathcal{P}}_{mj} = (-1)^m \frac{ h_N \cdots h_{m+1} }{ (h_j - h_N) \cdots 1 \cdots 1 \cdots (h_j - h_m) }.
\end{equation}

\begin{figure}[t]
    \centering   \includegraphics[width = 1\columnwidth]{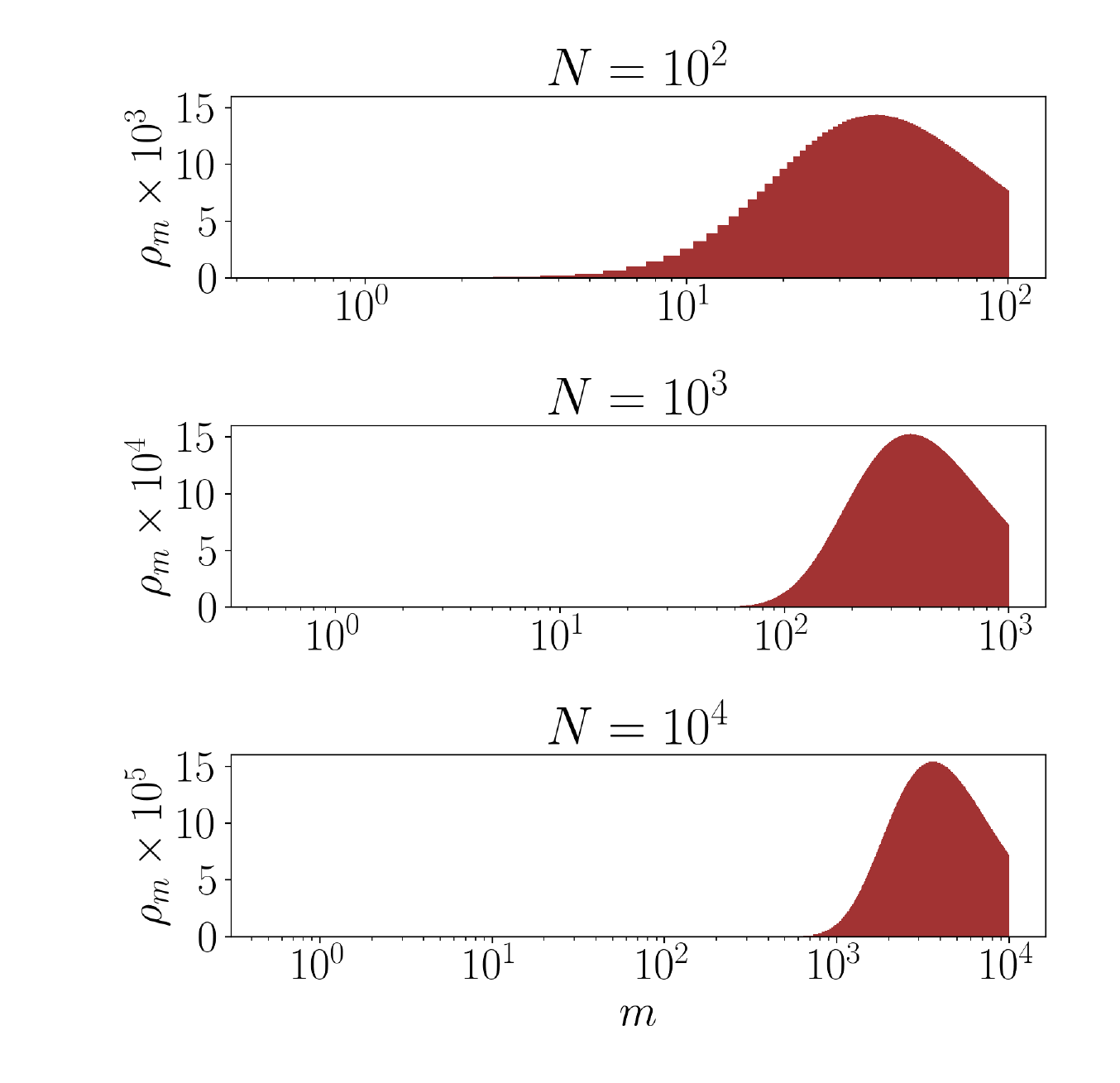}
    \caption{Distribution of Dicke states at the peak emission time, computed from the analytical expression in Eq.~\eqref{solution}. The distribution shows a more pronounced peak as the  particle number is increased. The emission rate is computed as $I=\Gamma\langle S^\dagger S\rangle(t)$ and it peaks around $t_\text{peak}\approx\ln{N}/(N\Gamma)$. We observe, that the distributions peak at $m=39$, $m=362$ and $m=3414$ respectively.}
    \label{fig:2}
\end{figure}

Together, the contributions from first-order and second-order poles fully determine the time-dependent populations \( \rho_m(t) \) for all \( m \). Notably, this approach bypasses the need to solve the master equation directly and remains computationally efficient even for large \( N \). In Fig.~\ref{fig:2}, we illustrate the population distributions at the peak of superradiant emission for increasing values of \( N \), demonstrating the scalability of the exact solution and its ability to capture the macroscopic structure of the radiative cascade.
\\

\noindent \textbf{Matrix formulation} --  
The analytical solution presented above can also be conveniently expressed in matrix form. This representation highlights the linear structure of the solution and is especially useful for numerical implementations.

We begin by collecting the diagonal elements of the density matrix—i.e., the Dicke state populations \( \rho_m(t) \)—into a column vector. Since the lowest state \( \rho_0(t) \) can always be recovered from the normalization condition \( \mathrm{Tr}[\rho(t)] = 1 \), we restrict attention to the remaining \( N \) populations \( \rho_1(t), \ldots, \rho_N(t) \). These form an \( N \times 1 \) vector, ordered from highest to lowest excitation number:
\[
\vec{\rho}(t) =
\begin{bmatrix}
\rho_N(t) \\
\rho_{N-1}(t) \\
\vdots \\
\rho_1(t)
\end{bmatrix}.
\]

Each component \( \rho_m(t) \) is known to be a linear combination of exponentials of the form \( e^{-h_j \Gamma t} \), where the \( h_j \) are the eigenvalues of the Lindbladian superoperator, as discussed earlier. Therefore, we can write:
\[
\vec{\rho}(t) = \mathbf{M} \cdot \vec{v}(t),
\]
where \( \vec{v}(t) \) is the vector of exponentials
\[
\vec{v}(t) =
\begin{bmatrix}
e^{-h_N \Gamma t} \\
e^{-h_{N-1} \Gamma t} \\
\vdots \\
e^{-h_1 \Gamma t}
\end{bmatrix},
\]
and \( \mathbf{M} \) is an \( N \times N \) matrix of time-independent coefficients.

This matrix \( \mathbf{M} \) is lower triangular due to the structure of the solution: higher Dicke states (larger \( m \)) only depend on exponentials involving equal or smaller \( h_j \), corresponding to transitions down the ladder. The entries of the matrix consist of two types of coefficients:
- \( \mathcal{P}_{mj} \): contributions from simple (non-degenerate) poles,
- \( \mathcal{R}_{mj} \): contributions from double (degenerate) poles that arise due to symmetric values \( h_m = h_{\bar{m}} \) with \( \bar{m} = N+1 - m \).

To make this structure explicit, we separate the matrix into two blocks:
- The upper block (for \( m > N/2 \)) contains only \( \mathcal{P}_{mj} \) terms.
- The lower block (for \( m \leq N/2 \)) includes both \( \mathcal{P}_{mj} \) and \( \mathcal{R}_{mj} \), where \( \mathcal{R}_{mj} \) may carry explicit time dependence due to the second-order residue structure.

The full matrix-vector product takes the form:
\begin{widetext}
\begin{equation}
\begin{bmatrix}
\rho_N (t)\\
\rho_{N-1} (t)\\
\vdots \\
\rho_{N/2+1}(t)\\
\hline
\rho_{N/2}(t)\\
\rho_{N/2-1}(t)\\
\vdots\\
\rho_1(t)
\end{bmatrix} =
\left[
\begin{array}{c c c c|c c c c}
\mathcal{P}_{N,N} & 0 &\ldots &0 & 0 & 0 &\ldots &0 \\
\mathcal{P}_{N-1,N} & \mathcal{P}_{N-1,N-1} &\ldots &0& 0 &0 &\ldots &0 \\
\vdots & \vdots & \vdots & \vdots  &\vdots  & \vdots & \vdots \\
\mathcal{P}_{N/2+1,N} & \mathcal{P}_{N/2+1,N-1}  &\ldots &\mathcal{P}_{N/2+1,N/2+1}& 0 &0 &\ldots &0 \\
\hline
\mathcal{P}_{N/2,N} & \mathcal{P}_{N/2,N-1} &\ldots &  \mathcal{R}_{N/2,N/2+1} & \mathcal{R}_{N/2,N/2} & 0 &\ldots &0 \\
\mathcal{P}_{N/2-1,N} & \mathcal{P}_{N/2-1,N-1} &\ldots &  \mathcal{R}_{N/2-1,N/2+1} & \mathcal{R}_{N/2-1,N/2} & \mathcal{R}_{N/2-1,N/2-1} &\ldots &0 \\
\vdots & \vdots & \vdots & \vdots  &\vdots  & \vdots & \vdots \\
\mathcal{R}_{1,N} & \mathcal{R}_{1,N-1}  &\ldots &\mathcal{R}_{1,N/2+1}& \mathcal{R}_{1,N/2}  &\mathcal{R}_{1,N/2-1} &\ldots &\mathcal{R}_{1,1} \\
\end{array}
\right]
\begin{bmatrix}
e^{-h_N \Gamma t}\\
e^{-h_{N-1} \Gamma t}\\
\vdots \\
e^{-h_{N/2+1} \Gamma t}\\
\hline
e^{-h_{N/2} \Gamma t}\\
e^{-h_{N/2-1} \Gamma t}\\
\vdots\\
e^{-h_1 \Gamma t}
\end{bmatrix}.
\end{equation}
\end{widetext}

\noindent
For implementation, the entries of the matrix \( \mathbf{M} \) can be compactly expressed using Heaviside step functions:
\[
M_{mj} =
\mathcal{P}_{mj} \, \theta_{j \geq m} \, \theta_{j > \bar{m}} +
\mathcal{R}_{mj} \, \theta_{j \geq m} \, \theta_{\bar{m} \geq j},
\]
where \( \bar{m} = N + 1 - m \). This expression ensures that:
- \( \mathcal{P}_{mj} \) contributes only for non-degenerate eigenvalues (simple poles), and
- \( \mathcal{R}_{mj} \) accounts for the degenerate cases (double poles) below the equator.\\
\indent Only the \( \mathcal{R}_{mj} \) coefficients carry explicit time dependence, while the \( \mathcal{P}_{mj} \) terms are constant in time. This matrix formulation thus provides an efficient and structured way to evaluate the full time-dependent solution for arbitrary system sizes \( N \).\\

\noindent \textbf{Starting in an arbitrary Dicke state} --  
We now extend our analytical framework to describe the time evolution of the system when it is initialized in an arbitrary Dicke state rather than the fully inverted one. That is, we consider an initial condition of the form:
\[
\rho(0) = \ket{m_0} \bra{m_0}, \quad \text{with } 1 \leq m_0 \leq N.
\]
This scenario corresponds to an ensemble that starts with only \( m_0 \) excitations rather than all \( N \). Since the system evolves strictly downward in excitation number due to collective decay, the accessible Hilbert space is now truncated: only Dicke states \( \ket{m} \) with \( 0 \leq m \leq m_0 \) will be populated during the evolution.
The derivation of the solution proceeds similarly to the fully excited case, but with the upper limit of the recursive structure now restricted to \( m_0 \). In particular, we follow the same residue-based construction as before, but restrict the set of poles to those between \( m \) and \( m_0 \). The result is:
\begin{equation} \label{any-state}
\rho_m(t) = \sum_{j = j_m}^{m_0} \text{Res}\left[ f_{m_0,m}(z,t) \right] \Big|_{z = h_j},
\end{equation}
where \( j_m = \max\{m, N/2 + 1\} \) for even \( N \), or \( j_m = \max\{m, (N+1)/2\} \) for odd \( N \), as before. The structure of the poles and degeneracies remains unchanged, except that the summation now stops at \( m_0 \), the highest allowed excitation number in the system's evolution.\\
\indent The corresponding generating function is now defined with respect to the new initial condition:
\begin{equation}
f_{m_0,m}(z,t) = (-1)^{m_0 + m} \frac{h_{m_0} \cdots h_{m+1}}{(z - h_{m_0}) \cdots (z - h_m)} e^{-z \Gamma t}.
\end{equation}
This function plays the same role as \( f_m(z,t) \) did in the fully excited case, generating residues corresponding to the eigenvalue contributions of the Lindbladian evolution. The expression accounts for both simple and double poles, and the same machinery for evaluating first- and second-order residues can be applied here.\\
\indent\textit{Generalization to mixed initial states} --  
The linearity of the Lindblad evolution makes it straightforward to generalize the result to any initial state that is diagonal in the Dicke basis. Consider a mixed state of the form:
\[
\rho(0) = \sum_{k=0}^N \rho_k(0) \ket{k} \bra{k}.
\]
Using linear superposition, the full solution is simply the weighted sum of individual evolutions from each pure component \( \ket{k} \):
\begin{equation}
\rho_m(t) = \sum_{k=m}^N \rho_k(0) \sum_{j = j_m}^{k} \text{Res}\left[ f_{k,m}(z,t) \right] \Big|_{z = h_j},
\end{equation}
where each \( f_{k,m}(z,t) \) is defined in analogy with the single-pure-state case. This result allows one to describe the evolution from any probabilistic mixture of Dicke states, making the analytical solution fully general within the symmetric subspace.\\
\indent This framework retains its computational efficiency regardless of the initial configuration, and remains exact for all system sizes and times.\\

\noindent \textbf{Conclusions and outlook} --  
We have presented a compact, exact, and fully analytical solution for the Dicke superradiance process, valid at all times and for arbitrary system sizes \( N \). The analysis is restricted to the fully symmetric subspace of \( N \) identical two-level quantum systems and yields closed-form expressions for the time-dependent populations of the Dicke states.\\
\indent Our approach rests on three key structural insights: 
\textit{(i)} the existence of recursive relations between neighboring Dicke states \( m \leftrightarrow m-1 \), which reflect the cascading nature of collective decay;  
\textit{(ii)} the realization that these recursions admit closed-form solutions; and  
\textit{(iii)} the identification of these solutions as residue sums of a single rational function in the complex plane.\\
\indent This last point connects the solution method to complex analysis: the time evolution can be expressed as a contour integral over a generating function, an interpretation developed further in Ref.~\cite{holzinger2025solvingdickesuperradianceanalytically}. Importantly, the same residue-based formulation extends naturally to the case of arbitrary initial Dicke states, allowing for broad applicability within permutation-symmetric quantum systems.\\
\noindent Several generalizations suggest themselves. A direct extension is the case of driven Dicke superradiance, where a coherent drive couples neighboring Dicke states and modifies the decay dynamics. This introduces coherent interference into the recursive structure and leads to richer dynamical behavior. Analytical work on this extension is currently underway.\\
\indent A second direction concerns open-system generalizations involving both collective decay and local incoherent pumping—relevant, for instance, to models of superradiant lasing~\cite{bohnet2012asteadystate}. In such cases, the dynamics include competition between individual pumping and collective emission, potentially giving rise to non-trivial steady states. The residue-based formalism introduced here may offer a path toward analytical treatment even in these more complex non-equilibrium settings.\\
\indent In summary, the analytical framework developed here—grounded in recursive relations and complex analysis—provides a tractable and transparent route to understanding collective emission processes, and opens a door to more general studies in dissipative many-body quantum optics.
\\

\noindent \textbf{Acknowledgments -} We acknowledge financial support from the Max Planck Society and the Deutsche Forschungsgemeinschaft (DFG, German Research Foundation) -- Project-ID 429529648 -- TRR 306 QuCoLiMa
(``Quantum Cooperativity of Light and Matter''). This research was funded in whole or in part by the Austrian Science Fund (FWF) [10.55776/COE1].



\bibliography{apssamp}

\clearpage
\pagebreak
\onecolumngrid

\section{APPENDIX}

\subsection{Solution of the recursive equation}

We provide here a detailed derivation of the time evolution of the diagonal elements of the density matrix as a power series expansion in time. We begin with the recursive relation derived from the Lindblad dynamics:
\begin{equation}
   \rho_{m-1}^{(j+1)} = -h_{m-1} \rho^{(j)}_{m-1} + h_{m} \rho^{(j)}_{m},
\end{equation}
which allows one to compute the $(j+1)$-th order term for state \( m-1 \) using the $j$-th order terms of the adjacent Dicke states \( m \) and \( m-1 \). At the top of the ladder (\( m = N \)), the recursion is initialized with \( \rho_N^{(0)} = 1 \) and \( \rho_N^{(j)} = (-h_N)^j \), reflecting the fact that the fully excited state decays exponentially at rate \( h_N \Gamma = N \Gamma \).

Proceeding downward through the Dicke ladder using the recursive formula, one can iteratively generate all coefficients \( \rho_m^{(j)} \) for fixed \( m \), and then sum the power series:
\begin{equation}
    \rho_m(t) = \sum_{j=0}^\infty \frac{(\Gamma t)^j}{j!} \rho_m^{(j)}.
\end{equation}
This summation leads to the closed-form expression
\begin{equation}
    \rho_{m}(t) =  \sum_{j=m}^N   \frac{e^{-h_j\Gamma t}}{\mathcal{D}_{m,j}},
\end{equation}
with
\begin{equation}
   \mathcal{D}_{m,j} =(-1)^m  \frac{\prod\limits_{k = m+1,\, k \neq j}^N (h_j - h_k)}{\prod\limits_{k = m+1}^N h_k}.
\end{equation}
Here, the product in the numerator excludes the term \( h_j - h_j \), which would otherwise lead to a divergence. This expression is valid for all \( m \geq N/2 + 1 \) (for even \( N \)) or \( m > (N+1)/2 \) (for odd \( N \)), where the spectrum \( h_m \) is non-degenerate. In this regime, each term in the sum corresponds to the residue of a simple pole in the generating function defined in Eq.~\eqref{eq:function}.

\subsection{Degenerate points below the equator}

For Dicke states below the equator, i.e., \( m \leq N/2 \) (or \( m < (N+1)/2 \) for odd \( N \)), the Dicke ladder spectrum exhibits degeneracies due to the symmetry \( h_j = h_{\bar{j}} \), where \( \bar{j} = N+1 - j \). These degeneracies give rise to second-order poles in the residue formulation of the solution.

To handle this, we treat the degenerate case using a limiting procedure. Let us assume
\[
h_j = h_{\bar{j}} + \epsilon
\]
and expand in small \( \epsilon \). One finds:
\begin{equation}
    \frac{e^{-h_{\bar{j}} \Gamma t}}{\mathcal{D}_{m, \bar{j}}} + \frac{e^{-h_j \Gamma t}}{\mathcal{D}_{m,j}} \approx \frac{e^{-h_j \Gamma t}}{\mathcal{A}_{m,j}} (\Gamma t - \mathcal{B}_{m,j}),
\end{equation}
where the denominators are expanded as
\begin{align}
\mathcal{A}_{m,m'} &=  (-1)^{m+1} \frac{h_{m'}^2}{h_m}
\prod_{\substack{j = 1 \\ j \neq m'}}^{m-1} \frac{h_j}{h_{m'} - h_j}
\prod_{\substack{j = m \\ j \neq m'}}^{N/2} \left( \frac{h_j}{h_{m'} - h_j} \right)^2, \\
\mathcal{B}_{m,m'} &= \sum_{\substack{j = 1 \\ j \neq m'}}^{m-1} \frac{1}{h_{m'} - h_j}
+ 2 \sum_{\substack{j = m \\ j \neq m'}}^{N/2} \frac{1}{h_{m'} - h_j}.
\end{align}
Although these expressions appear cumbersome, they can alternatively be obtained by calculating the residue of the generating function around a second-order pole, as outlined in the main text.

\subsection{Solution for odd \( N \)}

For odd system sizes \( N \), the spectrum features a unique non-degenerate point at \( m = (N+1)/2 \). The general residue-based solution remains valid, but this central point must be treated separately in the sum over residues. The full solution for arbitrary \( m \) reads:
\begin{align}
\rho_m(t) &= \sum_{j = m}^N \left[ f_m(z,t)(z - h_j) \right]_{z = h_j} \, \theta_{m \geq \frac{N+1}{2}} \nonumber \\
&+ \Bigg\{
\sum_{j = \bar{m}+1}^{N} \left[ f_m(z,t)(z - h_j) \right]_{z = h_j}
+ \left[ f_m(z,t)(z - h_{(N+1)/2}) \right]_{z = h_{(N+1)/2}} \nonumber \\
&\quad + \sum_{j = (N+1)/2}^{\bar{m}} \frac{d}{dz} \left[ f_m(z,t)(z - h_j)^2 \right]_{z = h_j}
\Bigg\} \theta_{m \leq \frac{N-1}{2}}.
\end{align}

Here, the second-order poles are treated via derivatives, while the non-degenerate center point contributes as a standard first-order pole. This formula recovers the general result given in Eq.~\eqref{solution}, properly modified to account for the symmetry breaking that occurs in the odd-\( N \) case.

\end{document}